# A minimal model of plasma membrane heterogeneity requires coupling cortical actin to criticality.

Benjamin B. Machta[1], Stefanos Papanikolaou[1], James P. Sethna[1], Sarah L. Veatch[2]
*Deparment of Physics (1) and Chemistry and Chemical Biology(2), Cornell University, Ithaca, NY*

## Abstract:

We present a minimal model of plasma membrane heterogeneity that combines criticality with connectivity to cortical cytoskeleton. Our model is motivated by recent observations of micron-sized critical fluctuations in plasma membrane vesicles that are detached from their cortical cytoskeleton. We incorporate criticality using a conserved order parameter Ising model coupled to a simple actin cytoskeleton interacting through point-like pinning sites. Using this minimal model, we recapitulate several experimental observations of plasma membrane raft heterogeneity. Small (r~20nm) and dynamic fluctuations at physiological temperatures arise from criticality. Including connectivity to cortical cytoskeleton disrupts large fluctuations, prevents macroscopic phase separation at low temperatures (T≤22°C), and provides a template for long lived fluctuations at physiological temperature (T=37°C). Cytoskeleton-stabilized fluctuations produce significant barriers to the diffusion of some membrane components in a manner that is weakly dependent on the number of pinning sites and strongly dependent on criticality. More generally, we demonstrate that critical fluctuations provide a physical mechanism to organize and spatially segregate membrane components by providing channels for interaction over large distances.

## Introduction

It is hypothesized that the plasma membranes of mammalian cells are heterogeneous over distances much larger than the nanometer size typical of their lipid and protein components (1-2). The physical origins and functional significance of this structure are contentious, and the hypothesis itself poses a thermodynamic puzzle: building an extended region rich in specific membrane components should cost a free energy proportional to the region's area due to the loss of entropy. The same structure potentially gains free energy proportional to its area by bringing together components which have lower interaction energies. Both of these effects are of the order $k_B T$ per lipid area, where $k_B$ is Boltzmann's constant and T is the temperature. Barring a remarkable cancellation, a domain with a size of 20nm would seem extremely unstable. Either the entropic contribution should win and the equilibrium structures should be much smaller, or energy should win, and the structures should be permanent and macroscopically phase-separated.

One way a cell could make stable domains with dimensions in the tens of nanometers is to carefully balance the entropic and energetic contributions, tuning near to a critical point (Fig. 1A). There, structures with characteristic sizes much larger than individual molecules emerge because the free energy required for their formation is near $k_B T$. When not exactly at the critical point these *critical fluctuations* are cut off at a size called the *correlation length*. In a simple system of two components, the two parameters that need tuning would typically be temperature and the ratio of the two components; in cell membranes at fixed temperature these two parameters could be the molar fraction of any components.

Remarkably, there is experimental evidence that plasma membranes of mammalian cells have compositions tuned near to a critical point at physiological temperatures. Giant plasma membrane vesicles (GPMVs) isolated from living cells appear homogenous to light microscopy at 37°C (310K), indicating that they are uniform at optical (250nm) length scales. However, below a critical temperature around 22°C (295K), these vesicles phase separate into two macroscopic fluid domains (3 ). Near the transition, GPMVs undergo equilibrium fluctuations that are visible at micron scales, in quantitative agreement with theoretical predictions of two dimensional (2D) criticality (4). One prediction that arises from this past GPMV study is that cell plasma membranes at physiological temperatures of 37°C (310K) reside only 5% above this critical point in the absolute Kelvin units natural to thermodynamics. This implies an experimentally extrapolated correlation length of roughly 20nm in GPMVs at 37°C (Fig. 1A). One major aim of this study is to demonstrate that criticality in plasma membranes, normally found only in carefully tuned laboratory environments, explains many experimental observations of membrane heterogeneity typically associated with 'lipid rafts.'

Plasma membrane vesicles differ from intact cell plasma membranes in many important ways; most notably, GPMVs lack connectivity to the cytoskeleton. In intact cells, the plasma membrane couples to cortical cytoskeleton through diverse and partially understood interactions (5-6). In addition to providing structural support, there is increasing evidence that the cytoskeleton plays a role in promoting lateral heterogeneity at the cell surface. It is widely speculated that connections to the cytoskeleton prevent the large-scale accumulation of membrane components into phase separated domains (3, 7-9), even under conditions where phase separation is readily observed when membrane-cytoskeleton coupling is disrupted; macroscopic phase separation is easily observed in cell-attached GPMVs even as the remaining intact plasma membrane remains homogeneous (Fig. 1B).

Here we explore a minimal model for an intact cell's nearly critical plasma membrane coupled to its cortical cytoskeleton by taking advantage of a remarkable property of nearly critical systems called *universality*. The shapes, sizes, and lifetimes of fluctuations depend only on the dimensionality of the system, the universality class, and the parameters that describe the relative proximity to the critical point (Fig. 1A). Universality enables us to make quantitative predictions about compositionally complex cell plasma membranes through simulations of much simpler model systems. We stress that though cell membranes are not exactly at a critical point at 37°C, they are tuned close enough to feel its universal features.

## Methods

All simulations use a square lattice Ising model with a conserved order parameter implemented in the standard way. A detailed description of all methods can be found in Supplementary Material and are summarized below. Temperatures are calibrated by setting the critical temperature of the Ising Model to 295K. All simulations are performed on periodic 400x400 arrays, with a pixel length corresponding to 1nm. Simulations performed to deduce static properties use non-local dynamics to decrease equilibration times. Dynamical simulations use Kawasaki dynamics supplemented with moves which swap like pixels at the same attempt frequency as unlike pixels. Simulation steps are converted into time using a conversion factor of approximately $D=1\mu m^2/s$. Correlation functions are normalized to one at spatial infinity.

We implement a cytoskeletal meshwork using a random, periodic, Voronoi construction to generate 'filaments' that have a width of one pixel (1nm). The pinning sites are chosen randomly along these lines with constant density, which is 0.4 except in Fig. 5 where it varies. A pixel sitting on a pinning site is constrained to be white. In diffusion experiments other white pixels are free to swap with pixels sitting on the pinning sites. Strongly coupled objects have infinitely strong interactions with their neighbors, forbidding any move which ends with a black pixel as a nearest neighbor to a strongly coupled white object or vice-versa.

## Results

**Overview of the Model.** We model the plasma membrane using a 2D Ising model as described in Materials and Methods. The Ising model, and modifications of this model, has previously been used to model the thermodynamic properties of purified bilayer membranes in the vicinity of the main chain transition temperature (10-11), where there is some evidence of critical behavior (12). In this work we have chosen to use the 2D Ising model to model membranes in the vicinity of their miscibility critical point because its behavior is in quantitative agreement with micron-sized fluctuations observed in isolated GPMVs and three component model membranes (4, 13) and because it represents the expected universality class for liquid-liquid phase separation (see Supplement).

In our model, membrane components such as lipids and proteins are represented as black or white pixels on a square lattice, where pixels of one type (e.g. white) correspond to components that tend to populate one membrane phase (liquid-ordered vs. liquid-disordered). We implement a conserved order parameter, meaning that the number of white (or black) pixels does not change with time. This model does not accurately reproduce the detailed arrangement of lipids and proteins that occur at very short distances (less than several nanometers), since their arrangement will depend strongly on detailed molecular interactions. In contrast, the Ising model is expected to produce a quantitative description of real membrane organization on larger distances where important fluctuations arise because the membrane is in close proximity to a miscibility critical point; its coarse-grained properties should agree, though its microscopic details are very different. Though theory and experiment suggest Ising criticality, we expect that our results would hold even if more exotic criticality turned out to be present in the system. They arise from a large correlation length and time, both of which are generically present in critical systems, and both of which have been directly observed in GPMVs (4).

In the Ising model, the critical point occurs when there are equal numbers of black and white pixels, with temperature tuned to the onset of phase separation. This corresponds to a membrane that has an equal surface area of liquid-ordered and liquid-disordered phase at the miscibility phase boundary. In most of this study, we assume that the plasma membrane has a critical composition, with a phase diagram similar to that shown in Fig. 1A. We also assume that the surface fraction of phases does not change substantially with temperature. Such a temperature dependence would lead to a tilt in the system's rectilinear diameter (Fig 1A). In simulations this would manifest itself as a temperature dependent change in the ratio of black to white pixels. Though experimental observations of GPMVs show nearly equal fractions of coexisting phases at temperatures well below $T_c$ ($T_c$-$T$~10°) we theoretically expect similar results even if this ratio were to show significant temperature dependence (see Supplement).

We generate a cortical cytoskeleton network from a Voronoi construction (see Materials and Methods), and fix individual pixels to be in particular state at random sites along these filaments. Pinned pixels are intended to represent membrane proteins with a preference for one membrane phase that are connected (either directly or indirectly) to both the plasma membrane and the underlying cytoskeleton. We pin pixels of the same type (white), presuming that the interaction with the cytoskeleton prefers one of the two low temperature phases. This description of cytoskeleton-membrane coupling is simplistic, but we expect it to capture the qualitative behavior as long as connections on average prefer one of the two lipid environments. Similar, though quantitatively different, results are obtained when both pixel types are pinned. We vary the linear density of pickets, along with temperature and composition. In plasma membrane vesicles, critical temperatures are typically near room temperature ($T_c$=22°C=295K) (4). We primarily investigate physiological temperature, $T$ = 37°C =310K=1.05 $T_c$ where $T_c$ is measured in Kelvin. We include simulations at physiological temperatures for membranes whose critical points are as low as 155K ($T$=2$T_c$), where correlations between components fall off within a few lipid diameters, in order to contrast behavior with possible cells whose membrane's composition is not close to a critical point.

**Phase separation is disrupted in the presence of cortical cytoskeleton**. Below $T_c$, in the absence of pinning, white and black pixels organize into domains that are half the size of the simulation box, indicating that the system is phase separated (Fig. 2A). In the presence of cytoskeletal coupling, components instead follow the template of the underlying meshwork (Fig. 2D). As a result, black and white pixels do not organize into domains that are larger than the characteristic size of the cytoskeletal corrals. If the average meshwork size is smaller than the optical resolution limit of light microscopy, as is the case in a variety of cell types (14-15), then our model predicts that intact cell plasma membranes would appear uniform even at temperatures where an isolated membrane would be phase separated. This provides an explanation for why intact cell plasma membranes generally appear uniform at low temperatures where macroscopic phase separation is readily observed in isolated GPMVs (Fig. 1B).

Our model is an example of a 2D Ising model with quenched (spatially fixed) random field disorder. A robust feature of the 2D Ising model (16) is that after the addition of any such disorder there is no macroscopic phase separation at any temperature (17). This holds for arbitrarily weak (or in our case sparse) random fields or if different pixel types are held at each pinning site. This lack of long range order depends on the fixed anchoring of the disorder; even slowly diffusing mobile proteins do not necessarily impede phase separation (18) (note that the GPMVs in Fig. 1B contain substantial protein content). The addition of mobile components could either raise or lower the transition temperature depending on the details of their interactions(19-20).

**Membrane fluctuations mirror the underlying cytoskeleton at physiological temperature.** In the absence of coupling to cytoskeleton, large composition fluctuations occur in simulations because the free energy cost of assembling a cluster with dimensions of a correlation length is roughly the thermal energy $k_BT$. At the critical temperature the correlation length is very large (in principle infinite, but cut off at the size of the simulation box, Fig. 2B), whereas at 1.05 $T_c$ the correlation length is roughly 20 lattice spacings (Fig. 2C). Since we equate one lattice spacing with 1nm, this is in agreement with the extrapolated correlation length in GPMVs at 37°C. When simulations are coupled to cortical cytoskeleton, the presence of the pinning sites disrupts the largest fluctuations (Fig. 2E,F). More strikingly, coupling to cytoskeleton entrains white pixel channels, leaving black pixel puddles in the center of each cytoskeletal corral (Fig. 2G). This occurs even though the cytoskeleton only interacts with the membrane at local pinning sites. The effect propagates over roughly a correlation length because the system is near a critical point.

We examine the extent of cytoskeleton-induced membrane heterogeneity in our model both near and far from criticality by averaging many snapshots like those shown in Fig. 2. Fig. 3A shows the time-averaged pixel value at each location in the image. Continuous and wide channels of white pixels follow the underlying filaments that make up the cytoskeleton. In simulations where the critical temperature is well below physiological temperatures (Fig. 3C, $T_c$=-120°C), the remaining channels of white spins are thinner, have gaps, and their contrast is dramatically reduced. This represents the case where the plasma membrane does not experience significant composition fluctuations at physiological temperatures, and highlights some of the effects of our model which arise because the membrane is tuned close to a critical point.

Our observations are quantified by evaluating pair correlation functions for the nearly critical case (Fig. 3B) and for the far-from-critical case (Fig. 3D). Pair auto-correlation functions, $G(r)$, measure the relative probability of finding a second pixel a distance r away from a given pixel of the same type. Pixels are significantly auto-correlated in simulations performed near criticality in the presence or absence of coupling to cortical cytoskeleton, and correlation functions have roughly the same shape (dashed lines in Fig. 3B). In simulations that are coupled to cytoskeleton, we also evaluated cross-correlation functions between membrane pinning sites and white pixels (solid lines in Fig. 3B,D). In simulations near the critical point there is an increased probability of finding a white pixel out to a distance of ~20nm (around a correlation length) from a pinning site (Fig. 3B). These long ranged correlations between the pinning sites and white pixels fill in gaps in the meshwork making the continuous channels shown in Fig. 3A. In simulations far from criticality ($T_c$=0.5T), both the autocorrelations and cytoskeleton cross-correlations fall off over a few nm due to the short range of the lipid mediated effective interactions (Fig. 3D). These correlation functions, especially the cytoskeleton-membrane component cross-correlations, are predictions of our model that can be measured experimentally, although their magnitude will depend on how strongly the specific component couples to the local membrane environment.

**Cytoskeleton stabilized membrane heterogeneity is long lived.** The lifetimes of typical fluctuations become increasingly long as the critical point is approached (21). In order to investigate this *critical slowing down* in our model, we implemented locally-conserved order parameter dynamics, where pixels may only exchange with their four closest neighbors. A microscopic diffusion constant of around D=1µm²/s was used to convert between simulation steps and seconds. Our dynamics assume that the composition is locally conserved while momentum is not conserved in the plane of the membrane due to interactions with the cytoskeleton and bulk fluid (see Supplement). The time-time correlation functions shown in Fig. 4A measure the probability of finding a pixel of the same type at the same location at a later time. Near the critical point with conserved order parameter dynamics, the correlation function decays with a characteristic time $\tau \sim \xi^z$, where the correlation length, $\xi$, is ~20nm at T=1.05$T_c$ and z is 3.75 (21). This means that even in the absence of coupling to cytoskeleton, fluctuations of ~20nm will on average live around 100ms which is a thousand times longer than the time for a single pixel to diffuse through this same distance, and roughly a million times longer than the time for a far from critical membrane to equilibrate. In the absence of cytoskeleton, correlations decay to the uncorrelated value of one at long times (~1s, Fig. 4A).

Time-time correlation functions for simulations conducted in the presence of cytoskeleton (red and green traces in Fig. 4B) are qualitatively similar for short times, but approach a value greater than one as t→∞. This occurs because the locations of cytoskeletal filaments are fixed in time. In a real membrane, these correlations will persist until the cytoskeleton rearranges, which we expect to be on the order of seconds to hours. This emergence of a slower time-scale of membrane organization correlated with the location of cortical cytoskeleton is a direct consequence of our model that could be measured experimentally.

**Membrane components undergo hop diffusion.** In addition to measuring the dynamics of membrane fluctuations, we also track the dynamics of individual components. Different species can partition into the low temperature membrane phases with non-universal partitioning coefficients, which might be stronger or weaker than the partitioning coefficients of our pixels. To explore a range of these, we track two types of objects. We track single pixels which interact with their neighbors with the same energies as the pixels that make up the bulk membrane. A possible example of this type of component might be a lipid present in high abundance in the plasma membrane. We also conduct separate simulations that contain a small fraction of components that couple more strongly to their local membrane environment, effectively forming extended cross structures with twice as many nearest neighbors and three times as many bonds to their local environment. Some examples of components which couple more strongly like this might be large transmembrane proteins which have contact interactions with a large number of nearest neighbors, or minority lipid species with extreme partitioning behavior such as long chain sphingomyelin lipids or polyunsaturated glycerphospholipids. The model that contains strongly coupled diffusing crosses has four components (black and white pixels and black and white crosses) but is expected to still be in the Ising universality class (22) (23). Representative tracks for strongly coupled diffusers are shown in Fig. 4B.

Diffusion is quantified by measuring mean squared displacements (MSDs) for a large number (1000) of tracked diffusers. In all cases, we find instantaneous diffusion constants somewhat lower than that imposed by the hop rate (1$\mu$m$^2$/sec). At longer times, MSD curves cross over to a second regime where objects appear to undergo slower effective diffusion, indicating that they are confined. Even in the absence of cytoskeletal coupling diffusers show slightly confined diffusion. Including cytoskeletal coupling leads to significant confinement of weakly coupled black diffusers (Fig. 4C), and even greater confinement of strongly coupled black diffusers (Fig. 4D). This occurs even though the cytoskeletal attachment sites have substantial gaps due to the entrainment of the white channels. The resulting black tracer diffusion behavior resembles the 'hop diffusion' reported for some plasma membrane components in living cells (14-15). Confinement effects are more pronounced for strongly coupled objects than for weakly coupled objects because there is a higher energy cost associated with having a strongly coupled object in an unfavorable local environment. In contrast, there is a significant probability that a single pixel will diffuse into a region rich in pixels of the other type because the energy cost to having four unlike neighbors is on the order of $k_B T$.

**Confinement depends strongly on criticality and weakly on pinning density.** Fig. 4 demonstrates the predictions of our model on the MSDs of diffusers for a specific set of parameters. From this data we can get two diffusion coefficients, one extracted from the value of the MSD at short times (500μs), and another from the value at long times (50ms). The ratio of these short- to long-time diffusion constants provides a measure of confinement that depends only weakly on details specific to the model (such as the imposed microscopic diffusion coefficient). In Fig. 5 we explore how this measure of confinement for strongly coupled black diffusers is modulated by distance to criticality and pinning density, and how it varies for both black and white objects as a function of composition and pinning density.

Fig. 5A demonstrates that, at a critical composition, strongly coupled black objects are confined as long as the average distance between pickets is smaller than the correlation length. In the nearly critical region (T ~ $T_c$) most relevant to biological membranes, sparse pinning sites are able to effectively block strongly coupled black diffusers. In contrast, pickets need to be extremely dense to produce confinement in membranes that are far from criticality in temperature (T >1.5$T_c$). As diffusion is *space-filling* in two dimensions, particles easily fit through openings without this long ranged effective force. White objects show little anomalous diffusion, even near the critical point, because they can diffuse along cytoskeletal channels.

We also examine how the diffusion of strongly coupled black and white objects is modulated by changing the total fraction of white and black pixels (Fig. 5B,C). The surface fraction of phases can be altered in plasma membranes by, for example, cholesterol depletion with methyl-β-cyclodextrin (24). In simulations, we probe a wide range of compositions by varying the fraction of black and white pixels at a constant temperature. Changing membrane composition modulates both the continuity of each pixel type and the correlation length of fluctuations (Fig. 1A). As before, we find that strongly coupled black objects are confined in nearly critical membranes. We more generally find that the confinement of strongly coupled black and white objects is primarily determined by the connectivity of their preferred phase. In the absence of coupling to cytoskeleton, there is a percolation-like transition when there are equal numbers of white and black pixels near $T_c$. (25) The presence of white pinning sites biases this transition towards larger fractions of black pixels. As a consequence, black objects have confined diffusion over a broad range of membrane compositions and pinning densities, while white objects are significantly confined only at low pinning densities and large fractions of black pixels. The magnitude of confinement arising from steric restrictions to diffusion is not expected to depend significantly on membrane temperature or composition, making this a robust prediction of our model.

## Discussion

In this study, we demonstrate that many reported properties of heterogeneity in cell plasma membranes are reproduced using a simple model that incorporates critical fluctuations and coupling to a fixed cortical cytoskeleton. Critical fluctuations that occur near the liquid-ordered/liquid-disordered miscibility critical point are inherently 'small, heterogeneous, highly dynamic' domains (26). In the absence of membrane-cytoskeleton coupling, the size, composition, and lifetime of fluctuations depends only on the relative proximity to the underlying critical point. In the presence of membrane-actin coupling, these are also governed by the dimensions and movement of the underlying cytoskeletal meshwork. We propose that relatively large (~20nm) and long lived (>10ms) fluid domains as are commonly described in the membrane literature are best understood as fluctuations arising from proximity to criticality.

Our model provides a simple explanation for why macroscopic domains are not readily observed in intact cell plasma membranes upon lowering temperature, even though macroscopic phase separation is routinely observed upon lowering temperature in vesicles made from purified lipids (27), cellular lipid extracts (28), isolated plasma membranes (3), and even in whole cells where plasma membranes are dissociated from cortical actin using detergents (29) or detergent-free methods (30). In our model, macroscopic phase separation is disrupted in intact cell membranes because the size of the underlying cytoskeleton meshwork puts an upper limit on the size of domains that can form in the membrane.

At physiological temperatures, in the single phase region above the critical temperature (4), our model yields more functionally relevant predictions. The presence of membrane-actin coupling leads to long lived fluctuations, whose lifetimes are determined by motion of the cytoskeleton. This coupling entrains channels of membrane components that favor cytoskeleton-membrane pinning sites, while compartmentalizing components that are associated with the other membrane state. If liquid-disordered preferring components tend to associate more closely with cytoskeleton connections (31), then we expect that liquid-ordered preferring 'raft' proteins and lipids would tend to be compartmentalized within actin-bound corrals. This situation most closely resembles the schematic depictions of 'lipid rafts' presented in the literature (1). Alternatively, if liquid-ordered preferring components tend to associate more strongly with cytoskeletal connections (32), then we expect to find liquid-disordered preferring 'non-raft' components more confined within actin lined corrals. We imagine that any given cell could potentially exhibit both behaviors, and that there could be significant variation within single cells and between cell types. We expect that removal of cytoskeleton would significantly alter the localization and dynamics of membrane components, as is frequently observed experimentally (33-35).

Our model also provides a plausible explanation for the diversity of diffusive behaviors seen for plasma-membrane-bound lipids and proteins. In the hop diffusion model presented in (14-15), plasma membranes proteins and lipids are confined within actin-lined corrals by physical barriers. We show that by including criticality, confinement can become more robust because entrained channels fill in gaps between neighboring pinning sites. Our model predicts that the confinement of membrane components can depend on their preference for the two membrane phases in addition to their physical size. This could have functional significance; a membrane-bound receptor could significantly alter its localization and mobility upon binding to a ligand if that event modulates its coupling to a particular membrane environment. Such allosteric modulation of a receptor's coupling could be a potent regulatory mechanism near criticality, possibly leading to spatial reorganization and functional outcomes.

Although it is not directly explored in this study, we also expect that larger membrane-bound objects will tend to couple more strongly to membrane phases based on the larger size of their interacting surface. Since each protein typically interacts with many lipids, lipid-mediated interactions between proteins can be much stronger and more interesting than a typical lipid-lipid or lipid-protein interaction (36). It is possible that the stronger coupling of larger objects is responsible for the significant changes observed for diffusing components upon cross-linking (37). If, in addition, cross-linked proteins or lipids become immobile, then they could stabilize membrane domains that are rich in membrane components that prefer the same phase, as is observed in patching experiments (32), and in cells plated on patterned surfaces (38). A similar mechanism could contribute to the accumulation of signaling proteins at sites of receptor cross-linking in mast cells or at the immune synapse (9, 39).

While the predictions of our model are in good agreement with many findings in the 'raft' literature, several results are not easily explained in this framework. The tight clustering of components as well as the well-defined stoichiometry of clustering reported in EM (34) and homo-FRET (33) studies is not explained by our model, since interaction energies that are large compared to $k_BT$ are required to maintain this organization. Also, we are not able to reproduce the spot-size dependent diffusive behavior of fluorescently tagged sphingolipids recently reported in living cells using STED microscopy (40). We could generate similar results if we were to allow for traced particles to experience transient pinning events, which have been observed for a variety of membrane proteins (35, 37).

Our model differs substantively from other explanations of membrane heterogeneity. Unlike micro-emulsion models, we do not require the presence of line-active components that localize on domain boundaries (41). We expect that the inclusion of additional components such as line active molecules, either as mobile or pinned components, would modulate critical temperatures as has been shown previously (20, 42). In our model, the long-range and dynamic nature of critical fluctuations do not require that additional energy be inserted into the system, as is needed in models that include membrane recycling to disrupt macroscopic phase separation (43). We expect that recycling of membrane components will be important to understand the behavior at times on the order of membrane turn-over rates (min-hours) (44), which are significantly longer than those explored in this study. Our model also assumes that criticality arises from being close to a miscibility critical point that involves only liquid phases, and not a critical point postulated to be present near a transition to a gel phase (11-12).

In conclusion, we have presented a minimal model to explain the thermodynamic basis of heterogeneity in living cell membranes. Our model proposes that critical fluctuations, modulated by connectivity to cortical cytoskeleton, are both necessary and sufficient to explain the phenomena associated with 10-100nm fluid domains commonly described in the raft literature. In this new description of 'lipid rafts,' one major role of lipid mediated heterogeneity is to provide effective long range forces between membrane proteins that govern their organization and dynamics. Importantly, the cell could tune effective interactions between proteins by modulating overall membrane composition or by specifically altering the partitioning behavior of individual proteins. In this way, membrane heterogeneity can have direct implications on a wide range of cell functions.

## Acknowledgements

We thank Barbara Baird, David Holowka, Klaus Gawrisch, and Harden McConnell for helpful discussions and thoughtful reading of the manuscript. We acknowledge support from DOE office of Basic Energy Sciences (SP: DE-FG02-07ER46393), the NSF (JPS: DMR-0705167), the NIH (SLV: K99GM087810), and the Miller Independent Scientist Program of Cornell's Department of Chemistry and Chemical Biology (SLV).

# Figures

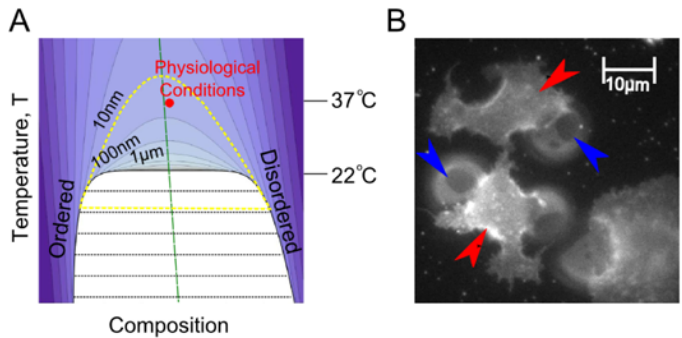

Fig. 1: <u>Ising criticality in the Plasma Membrane.</u> (A) Our model assumes that cell plasma membranes are tuned to the proximity of a 2-D Ising critical point with a critical temperature near room temperature (22°C) (4). Contours show regions of constant correlation length. The shapes of the contours are identical for any system in the 2-D Ising universality class except for the slope of the rectilinear diameter (green line's tilt, see supplement), which describes how the fraction of phases changes with temperature. The overall temperature scale of the contours comes from experiments in GPMVs. Most simulations are conducted at the red point which is hypothesized to represent physiological conditions. (B) Below the critical temperature, intact plasma membranes on living cells appear uniform at optical length scales (red arrows), while attached plasma membrane vesicles are macroscopically phase separated (blue arrows point to phase boundaries). Detailed methods for (A) and (B) are provided in Supplementary Material.

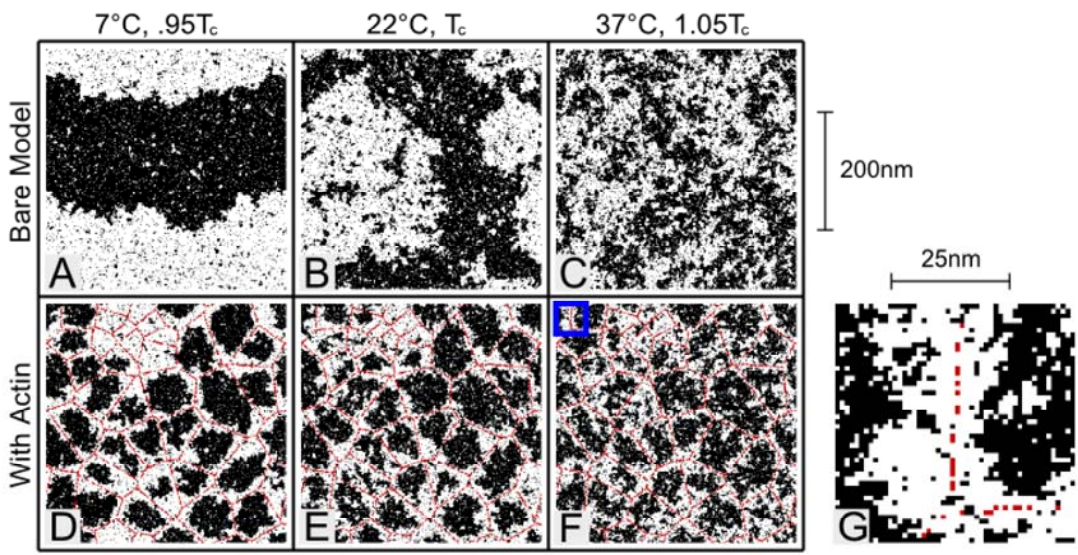

Fig. 2: <u>Membrane lateral heterogeneity is modulated by coupling to cortical cytoskeleton</u>. Ising model simulations were conducted over a range of temperatures in the absence (A-C) and presence (D-F) of coupling to a cortical cytoskeleton meshwork. Red sites indicate locations where pixels are fixed to be white. Below $T_c$ the bare Ising model is phase separated (A). Long range order is disrupted when the model is coupled to cortical cytoskeleton (D) because the structure is cut off at the length of the cytoskeletal corrals. At $T_c$ the bare model has structure at all length scales (B), whereas coupling to cytoskeleton cuts off the largest fluctuations (E). Above $T_c$, composition fluctuations that form in the bare Ising model (C) tend to localize along cytoskeletal filaments in the presence of coupling (F). (G) A higher magnification image (from boxed region in F) highlights that the cytoskeleton-preferring white phase forms channels around filaments with a width given roughly by the correlation length (20nm). The linear pinning density is 0.4 nm$^{-1}$.

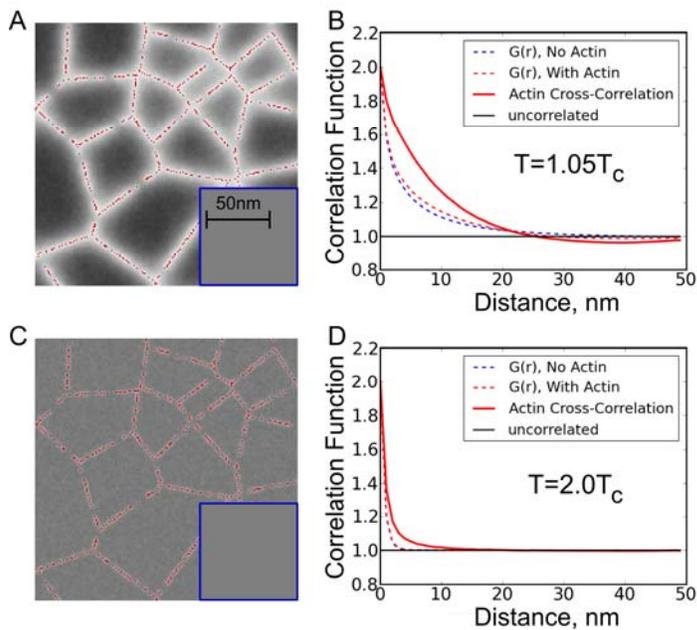

Fig. 3:  Coupling to cytoskeleton acts to entrain channels of white pixels over filaments, leaving pools of black pixels within cytoskeletal corrals. (A) The time averaged density of white pixels is correlated with the position of the cytoskeleton at 37°C (1.05 $T_C$). In the absence of cytoskeletal coupling (inset) the average density is trivially uniform. (B) The spatial correlation function, G(r), is not significantly altered by the presence of cytoskeletal coupling (compare blue and red dashed lines). In each case G(r) decays over a correlation length of roughly 20nm. In addition, the spatial cross-correlation function between white pixels and pinning sites (solid line) indicates that long range correlations extend over roughly one correlation length. (C) In a hypothetical 'generic' plasma membrane not tuned to the proximity of a critical point at 37°C, but instead with a critical temperature of -120°C, the channels gathered by the cytoskeleton are much thinner, and their contrast is diminished. (D) All of the corresponding correlation functions decay over a much shorter distance.

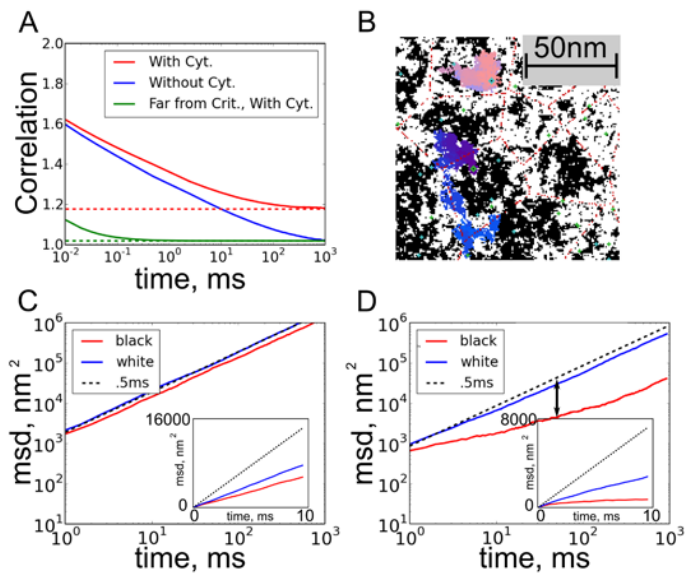

Fig. 4: <u>Membrane dynamics and component diffusion are sensitive to criticality and connectivity to cortical cytoskeleton</u>. (A). Near the critical point ($T_c$=0.95T), time-time correlation functions for membranes without coupling decay slowly, and become uncorrelated after roughly 1s (blue). In the presence of coupling to cortical cytoskeleton, fluctuations remain correlated even after long times (dashed red line at 'infinite' times). By contrast, systems that are far from critical (green line, $T_c$=0.5T) are uncorrelated after a fraction of a millisecond and coupling them to the cytoskeleton makes them decay to a small non-zero value (dashed green line). (B) Dynamics (at $T_c$=0.95T) are also measured by tracking components through simulation time-steps. Tracks for single black (pink) and white (blue) strongly coupled diffusers are shown (see text). (C-D) Mean squared displacements (MSDs) are calculated from many traces and indicate that (C) weakly coupled black lipids are slightly confined while (D) more strongly coupled black crosses are more strongly confined. Freely diffusing particles have MSDs that are linear in time (slope 1 or linear in inset). We quantify the confinement by the ratio of $D_{500\mu s}$ to $D_{50ms}$, whose log is the length of the double sided arrow.

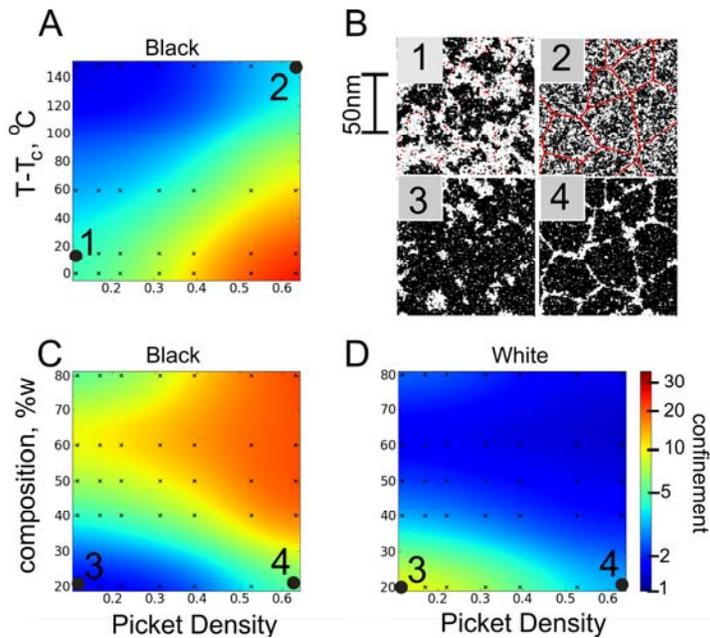

Fig. 5: <u>Confined diffusion depends upon criticality and the linear density of pinning sites</u>. (A) The ratio of $D_{500\mu s}$ to $D_{50ms}$ obtained from MSD curves like those shown in Fig. 4 are used to quantify the confinement of black crosses as a function of temperature and picket density. Near criticality, very weak pinning sites induce a large amount of confinement, whereas far from criticality even dense pinning leads to only slightly confined diffusion. (B) Representative simulation snapshots 1 and 2 have similar levels of confinement and have parameters indicated in part A. (C, D) The ratio of $D_{500\mu s}$ to $D_{50ms}$ is plotted as a function of composition and picket density plot at 37°C ($1.05T_c$) for both black (C) and white (D) traced crosses. When composition is varied, whichever of the two types is disconnected diffuses with more confinement (part B, images 3 and 4). The colored surface is a smoothed interpolation of the values from the black data points.

# Supplemental Information

## 1. Introduction:

Sections 1-4 of this supplement serve to give more complete details of the methods used in the simulations. Sections 5-8 discuss theoretical details which underlie our choice of model, discussing in more detail its benefits and limitations. In sections 2 and 3 we discuss the construction of figure 1. In section 4 we describe our simulations in sufficient detail that a reader could reproduce figures 2-5 in the main text. In section 5 we explain why we may, without significant effect, neglect the tilt of the rectilinear diameter in figure 1A of the main text. In section 6 we give arguments justifying our use of the Ising model to describe cell plasma membranes. In section 7, we justify our use of 'model B', Kawasaki dynamics. In section 8 we discuss how our results depend on the relative viscosities in the two phases.

## 2. Calculation of Correlation Length Contours in Figure 1A:

Following Combescot et al. [1] the correlation length as a function of thermodynamic parameters has a simple form in a sort of 'polar' coordinates[2-3]. A coordinate transformation takes these to the more familiar axes of reduced temperature and magnetization. We change coordinates a second time by introducing a non-zero arbitrary tilt, allowing for a non-universal correction which arises if a change in real temperature in a membrane corresponds to a change in both reduced temperature and magnetization in the Ising model. In the critical region this tilt is captured in a single parameter, continuous through the critical point, called the slope of the *rectilinear diameter* [4], which is non-universal and so may be different for different systems in the same universality class. Though it has not been quantified, in GPMVs it is within experimental bounds of zero as can be seen in the supplemental videos of [5]. It could theoretically be measured by looking at the change in the surface area of bright to dark regions as temperature is lowered below the critical point. Though experiments cannot differentiate the true tilt from zero, we include a small tilt here of 0.1 to stress that we do not expect it to be exactly zero. The contours all multiply a non-universal 'natural length scale' which is taken from experiments in plasma membrane vesicles [5]. Finally, we note that GPMVs in [5] have a spread in their critical temperatures of around 10°C, which corresponds to a spread in their reduced temperatures at 37°C of about .03. Though it has not been quantified, we expect there is also some spread in their effective magnetizations. Together, we expect that there is some variation from cell to cell both vertically and horizontally in the exact placement of the red dot corresponding to physiological conditions. We choose an average value of $T_c$=22°C at critical composition for this figure and most of the results in the paper. Ignoring a possible deviation in the magnetization is justified in section 5 below.

## 3. Preparation of cell attached plasma membrane vesicles in figure 1B:

Cell attached plasma membrane vesicles are prepared as described previously [5-6]. Briefly, RBL-2H3 mast cells are plated sparsely in a MatTek well (MatTek Corp.

Ashland, MA) overnight.  Cells are then incubated for 1h at 37°C in the presence of active bleb buffer (2 mM CaCl$_2$ /10 mM Hepes/0.15 M NaCl, 25 mM HCHO, 2 mM DTT, pH 7.4).  Cells and attached vesicles are then labeled with DiIC12 (Invitrogen.Eugine, OR) dissolved in methanol for 5min prior to viewing on an inverted microscope (DM-IRB; Leica Microsystems, Bannockburn, IL) at room temperature.  The image was taken using an EMCCD camera (iXon 897; Andor, Belfast U.K.). Under these conditions, attached blebs contain coexisting liquid-disordered (bright) and liquid-ordered (dark) phases.

### 4.  Simulation details, acceptance criterion and equilibration procedures:

All simulations were run on a 400x400 bi-periodic square lattice with spin variables living on the squares (S$_i$=±1, white and black pixels respectively). In all cases the standard Ising Hamiltonian is used, given by $H = -\sum_{\{i,j\}} S_i S_j$ with summation over the four nearest neighbors.  We use a conversion factor from lattice constant to real distance of 1nm.   Temperatures are converted to Kelvin (and by extension Celsius) by equating the exact critical temperature given by the Onsager solution on the square lattice,

$$T_c = \frac{2}{\log(1+\sqrt{2})}$$ with 295K so that $T_{sim} = 0.00769 T_{real}$, where $T_{real}$ is in Kelvin.  In a

Monte-Carlo 'sweep' 160,000 (400$^2$) pairs of pixels are proposed (on average each pixel is proposed to exchange twice).  We use Metropolis spin exchanges; each pair is exchanged or not so as to satisfy detailed balance[7]. If the resulting configuration is lower in energy, the exchange is always accepted.  If the energy is raised, the exchange is accepted stochastically with probability $\exp(-\beta \Delta H)$ where β is the inverse temperature and $\Delta H$ is the change in energy between initial and final states.  Sites occupied by pickets are taken to have an infinite field so that exchanges which propose to move a black pixel onto a picket are always rejected.  Where appropriate, 'strongly coupled' tracers have an infinite coupling to other like pixels, so that any move which ends with such a strongly coupled tracer touching an unlike pixel is always rejected. This effectively converts them into cross shaped objects (though with overlap allowed), with three times as many bonds to their local environment and twice as many neighbors.

Two types of dynamics are employed (any dynamics which satisfy detailed balance will lead to the same equilibrium ensemble of configurations).  When rapid equilibration is desired we employ nonlocal moves where each of a pairs of spins are randomly chosen from all sites on the lattice.  To simulate real time we use Kawasaki dynamics where we randomly choose a spin, and then randomly choose one of its four nearest neighbors to exchange with.

Equilibration is very rapid using the nonlocal dynamics, where z is near 2.  We always equilibrate for 100,000 sweeps using nonlocal moves starting from a distribution which observes the random field constraint but which is otherwise random.  100,000 sweeps is many times longer than the decay time of the slowest decaying system used here (the decay time is around 1000 sweeps for the pinning density in Fig. 2-4 at 1.05T$_c$.  as can be seen by eye looking at snapshots at subsequent times, or quantitatively as the decay time of time-time correlations).  For simulations with strong tracers we first equilibrate without tracers.  We then add them randomly, run an additional 100,000 iterations to equilibrate, and then start our dynamic simulations.

 In figure 2 no dynamics are required as only a snapshot is given.  In figure 3, the time averaged spatial correlation figures are averaged from 1000 snapshots each separated by 1000 sweeps of the non-local dynamics.  The auto-correlation functions in figure 3 are produced in the standard way. We first Fourier transform the spin configuration.  We then square this to get the static structure factor S(k). We then perform an inverse Fourier transform on S(k) and radially average the result to get g(r) in a normalization which goes to 0 at infinity.  To convert to the normalization used here, $G(r) = \dfrac{<P_{+1}(R)P_{+1}(R+r)>}{<P_{+1}(R)P_{+1}(\infty)>}$

(where R is averaged over the entire lattice and where $P_{+1}(x)$ is the probability of an up spin at position $x$) we add one to these (since all of our correlation functions are plotted at m=0).  We assume that the lattice sits on an infinite periodic plane so that values at $\infty$ take the mean value of the system.

To produce the cross-correlation curves we follow the same procedure except that we replace the square of the Fourier transform with the real part of the product of the Fourier transform of the pixel configuration and the Fourier transform of the random field configuration.  This leads to  $G(r) = \dfrac{<Pcyt(R)P_{+1}(R+r)>}{<Pcyt(R)P_{+1}(\infty)>}$  where $P_{cyt}(x)$ is the probability of finding a cytoskeletal pixel at position $x$.

For time-time correlations shown in Fig 4a we take the dot product of every pixel in the simulation's value ($\pm1$) at time t with itself at a later time t+$\Delta$t.  We average this over all pixels and all times t<$t_{max}$-$\Delta$t, with $t_{max}$=5,000,000 sweeps and add one to the value for consistency with our normalization (in supplemental fig S1 two different realizations of this procedure are shown to give the reader an idea of the expected error).  To produce the dashed lines which correspond to the asymptotic values of the correlation functions in the presence of the random field we take the average value of the square of the mean field pixel values from the configuration calculated from the non-local dynamics (which are identical and faster to equilibrate, displayed in Fig3A,C) and add one to this value.

To convert from Monte-Carlo time to real time, we use a microscopic diffusion constant of 1$\mu m^2$/s which is in the middle of the range cited for membrane bound lipids and proteins (though this range spans some 2 orders of magnitude)[8-10].  In a Monte-Carlo sweep, each pixel is proposed to swap twice on average.  If all swaps were accepted this would lead to a mean squared displacement  $<x^2>$ = 2 $d^2$/sweep where d is the lattice spacing.  With our value of d=1nm, we find that if we associate one sweep with .5$\mu$s we arrive at the desired D=1$\mu m^2$/s in the formula for diffusion $<x^2>$=4Dt. As some moves are rejected, the effective diffusion constant even at arbitrarily short times is actually somewhat lower than this for traced diffusing pixels and slower still for our more strongly coupled diffusers.

To calculate mean squared displacements we trace 1000 particles which diffuse on an infinite plane whose configuration is periodic with period 400.  Whenever a particle moves through a boundary in the $\pm$ x direction (for example), its new position for the purpose of mean squared displacement calculation is changed by $\pm$400.  This allows us to keep track of tracers which may diffuse off of the edge of the periodic simulation.  We average the mean squared displacements over all traced particles.

To produce the contour plots in fig 5 we extrapolate and smooth between the points where simulations are conducted by replacing each point's value by an average over all simulated points weighted by $\exp(-d^2/d_0^2)$ where $d_0 = .1$ in temperature, pinning density and magnetization.

## 5. Irrelevance of a tilted rectilinear diameter to lowest order:

As discussed in 'Calculation of correlation length contours' a change in real temperature near a critical point changes both of the corresponding Ising variables of reduced temperature and magnetization. Here we show that such a tilt introduces a subdominant correction to the critical properties. In particular it does not affect the singular behavior of the correlation length

In the scaling regime, a temperature change corresponds to a change in the reduced temperature by an amount $\Delta t$ and a change in the magnetization by an amount $\Delta m = a\Delta t$ where a is the tilt of the rectilinear diameter[4]. We here consider a model in the scaling regime whose correlation length is given by $\xi(t,m) = t^{-\nu}F(m^{\nu/\beta}/t)$ where $F$ is a universal function. We note that $F$ is not singular at zero, since along the $m = 0$ axis the correct scaling is given by $\xi(t,0) \sim t^{-\nu}$. We now set $m = at$ which corresponds to a membrane with critical composition taken to a temperature $T = (1+t)T_c$. This gives $\xi(t,m) = \xi(t,at) = t^{-\nu}F(a^{\nu/\beta}t^{\nu/\beta-1})$. Because $\nu/\beta = 8 > 1$, the argument of F is not singular near t=0. As F itself is not singular there, the scaling of the correlation length $\xi(t,m)$ at critical composition has singular behavior $\xi(t,0) \sim t^{-\nu}$ which is independent of the tilt of the rectilinear diameter. (In addition to the "analytic correction to scaling" represented by the tilting of the rectilinear diameter proportional to $T_c - T$, there is also a singular correction to scaling proportional to $(T_c - T)^{1-\alpha}$. Since $\alpha$=0 for the 2D Ising model, our argument above applies without modification; this correction too is subdominant. See [3] and [11] for details.) This means that the critical properties of an Ising system at critical composition but slightly away from the critical temperature are dominated by its effective reduced temperature, with its effective magnetization coming in only at a higher order in the distance from the critical point. This calculation quantifies what can be seen in Fig 1A of the paper; Near the critical point the contours are broad and flat showing a larger dependence on the vertical 'reduced temperature' direction than on the horizontal 'magnetization' direction.

## 6. Possible universality classes:

Here we give arguments to support our use of the 2-D Ising universality class to model the critical point seen in GPMVs [5]. We explain the theoretical motivation for expecting Ising criticality, discuss experimental evidence for it and argue that two other possible universality classes are unlikely to describe cellular membranes.

The Ising Universality class is expected for any system with a *scalar order parameter*, a single number which describes the system at larger length scales. In three dimensions, the Ising model has been shown to quantitatively describe an enormous array of liquid-gas and liquid-liquid critical points [3] mostly involving small molecules, but also including more exotic liquid phases containing polymer blends [12]. In two dimensions there are fewer examples, but the critical phenomenon seen in three component model membranes near fluid-fluid critical points are in this class [13]. In each of these two and three dimensional examples the components (or densities) of the two low temperature phases acts as the order parameter, which is therefore a scalar. Plasma membrane vesicles are certainly more complicated than the simple systems described above at the microscopic level, but as they phase separate into two domains with different compositions below $T_c$, the composition difference between these phases remains a good scalar order parameter. As such, the theoretical expectation is that they should also be described by 2-D Ising Universality, which is in agreement with [5].

We cannot exhaustively dismiss other possible universality classes at present, though several cases can be ruled out. The q-state Potts universality class generalizes the Ising model (q=2) to the critical point of a system which separates into q distinct liquids below $T_c$ [14]. For q=3 and 4 there are 2-D critical points, while for q larger, there are only abrupt, first order transitions. These models are ruled out quite simply by the GPMV experiments. The q-state Potts model predicts that below $T_c$ a Potts critical membrane should phase separate into q domains of approximately equal area. To our knowledge no more than two coexisting macroscopic liquid phases have been observed in any membrane system. We note as an aside that Potts models with q>2 would be dramatically harder for a cell to tune towards. The Ising critical point contains two parameters which must be tuned. At fixed temperature we must tune the ratio of the two phases below $T_c$ and their interaction energies. The 3-state Potts model contains five parameters that need tuning [14]. We can think of these as two area ratios (A:B and B:C) as well as three interaction energies (A with B, B with C and A with C).

Another possibility is that the membrane might display tri-critical Ising behavior. This occurs as an Ising model is tuned (along a third dimension in parameter space) to a boundary beyond which it becomes an abrupt first order transition. This model would only require tuning one additional parameter making it at first pass more appealing. However, it predicts a value for the critical exponent of $\nu=5/9$ [14], in contrast to $\nu=1$ for the Ising model, which is not consistent with existing experimental data [5].

Although we are not able to conclusively demonstrate that the universality class is Ising at this time, we emphasize that our results should hold even if the universality class turns out to be more exotic. The qualitative features of our findings come about due to features which have been conclusively demonstrated in GPMV experiments [5]: A correlation length which becomes large near the critical point, and dynamics which become slow near the critical point. Although it is outside of the scope of this work, we also note that there has been significant theoretical progress towards a complete categorization of possible universality classes permissible in 2-D [14]. Though there are infinitely many, they are all indexed by a unique number between zero and one, the central charge (which is 0.5 for the Ising model). It would be an interesting project to see which if any others might be consistent with membrane experiments.

## 7. Dynamic universality and motivation for model B:

The Ising universality class defines the coarse grained static correlation functions of our system. However, different systems in the same universality class can display different dynamics even in the scaling regime. These in turn fall into different dynamic scaling universality classes[15] which are determined by which quantities are conserved by the dynamics. In our case, we argue that the order parameter (or composition) is locally conserved, while momentum is not, as the fixed cytoskeleton breaks translational invariance, and may exchange momentum with the membrane. With the order parameter conserved and momentum not, we expect model B[15]. The Kawasaki dynamics we implement here are also in this class[16].

Membranes are expected to have a conserved order parameter for the times relevant to this study. The two low temperature phases contain different concentrations of various components. For a region's order parameter to change, components must physically move into it from a neighboring region. Although components are found with some probability in each low temperature phase (and are able to move between them), this is not qualitatively different from the Ising model where white pixels and even white pixel clusters are found in the low temperature black pixel phase. At longer times we expect processes like trafficking of lipids to change the order parameter[17].

We also expect that in intact cell plasma membranes momentum will not be conserved at relevant lengths, leading to model B (rather than model H, which arises when momentum is also conserved). There is an emerging theoretical picture for the expected dynamics of model membranes near an Ising critical point, immersed in water. In two dimensions the usual Stokes-Einstein relations which predict the microscopic diffusion constant as a function of the diffuser's size and the viscosity of the surrounding fluid cannot be easily applied [18] essentially because energy is not locally dissipated. For large inclusions this means that even an arbitrarily small viscosity of the surrounding fluid enters into the microscopic diffusion rate. It also means that this diffusion constant depends only logarithmically on the size of the diffuser, crossing over to a rate determined by the 3-D viscosity (which scales as 1/r) for large enough r [19]. For lipids in a bilayer membrane, the picture is somewhat simpler. Such lipids show an enormous temperature dependence in their microscopic diffusion rates [10] consistent with an energy barrier of 20-30 $K_BT$. These rates are approximately two orders of magnitude faster than the rates predicted by the hydrodynamic diffusion constant extrapolated from the movement of micron sized diffusers [19] in similar liquid environments. This suggests the following picture [20]: rather than their motion being dominated by hydrodynamic flow, lipids sit in deep potential wells in the membrane. Their microscopic diffusion rate is set by the likelihood of thermally hopping into the next potential well – model B, rather than model H, governs the particle diffusion rate. Thus membranes are similar to liquids above the glass transition, where self-diffusion (mediated by swapping particles) is much faster than bulk diffusion. (In a crystal, the latter would be zero.)

Even though particle diffusion is dominated by model B, we must address also the evolution of the order parameter field. Hydrodynamic flows are more effective at "stirring" the order parameter field than particle exchanges. This is reflected in the dynamical critical exponent z, which is 3.75 for model B and around 2 for model H in two dimensions [15, 21]. Roughly, correlation times at a length-scale L scale as $(L/L_0)^z$ (critical slowing-down). Since at the lipid length scale of one nm the time scales for hopping and hydrodynamic rates differ by two orders of magnitude, this suggests that the hydrodynamic effects will become competitive with the model B dynamics at roughly 10nm (where $D_{lipid}(L/L_0)^{3.75} \sim D_{hydro}(L/L_0)^2$), which is roughly the equilibrium correlation length where both power laws stop applying. In the absence of a cytoskeleton and at the critical temperature, the hydrodynamic diffusion will dominate in a range of lengths between this crossover and a crossover to a modified three-dimensional model H dynamics (when the low viscosity of the surrounding water becomes relevant, at around 1000nm) [21]. The rigid cytoskeleton will act as a fixed boundary condition at a significantly shorter length scale in our model, suppressing hydrodynamic flows entirely while permeable to hopping diffusion. (The cytoskeleton should be particularly effective at suppressing the logarithmic correlations in the 2D hydrodynamics.) Hence our model B dynamics not only dominates the diffusion of small particles, it also should dominate the dynamics of the order parameter field except for small corrections in a range intermediate between the correlation length and the cytoskeleton confinement scale. We finally note that coupling to a cytoplasm with many rigid objects nearby may lead to even more suppression of bulk flow in the membrane as discussed in [22]

## 8. Effects of different viscosities in the two liquid environments:

The $l_0$ and $l_d$ phases represented by our white and black pixels have viscosities which differ by a factor of around 4 though in some cases up to a factor of 10 [9-10, 23]. A similar range is expected in the diffusion constant difference seen between lipids in the two phases. In most of the manuscript we ignore the consequences of this, but we discuss its implications here. The dynamics of the order parameter are relatively unaffected by this as order parameter changes always take place at the interface and so have a single rate which is presumably somewhere in between the rates predicted by the individual viscosities.

Traced particle diffusion, however, can be affected by this viscosity difference. We consider the case relevant to our dynamics; a particle which mostly resides in one of the two phases, but which may need to cross through the other phase to diffuse long distances. The microscopic diffusion constant will be an average of the diffusion constants in the two phases weighted by the time spent in each phase. For us it will be given primarily by the diffusion constant in the phase in which it usually resides.

To travel large distances these particles potentially needs to hop over barriers of the alternative phase, which leads to the confinement seen in our simulations. We separate this process into an attempt rate at crossing and a success probability. The total amount of time spent in the unfavorable region, as well as the success probability are determined by static energetic considerations; they does not depend on the relative viscosities. The attempt rate, however, must depend on the ratio of the diffusion constants in the two liquids. In particular, to satisfy detailed balance it must go as the ratio of the viscosity in the usual fluid environment to that in the barrier environment. As such, particles which mostly live in the low viscosity environment make fewer attempts at crossing the barrier, while those which live mostly in the higher viscosity environment make more. This can lead to some additional confinement for particles which live in the lower viscosity phase. The extra confinement is bounded by the ratio of the two viscosities.

To demonstrate these theoretical predictions we run simulations where the diffusion rates in the two liquid environments are different by a factor of four, mimicking a factor of four change in the viscosity. We implement this by trading like particles in the lower viscosity liquid with a rate four times that with which unlike particles and particles in the high viscosity liquid are traded.  We run simulations where either the white pixels or the black ones have a higher viscosity.  We then plot MSDs vs time (fig S2), for the cases where both liquids are equivalent, and where the given particle is either a component of the lower or higher viscosity liquid.  In each case the y-axis is normalized to 1 at t=1ms (in the equal viscosity case), and we compress the x-axis for the high viscosity case so that the frequency of moves per unit 'time' on the x-axis is the same.  We also plot a dashed line corresponding to a lack of any confinement.  We note that in our simulations this consequences of this effect are fairly small.

# Supplemental Figures

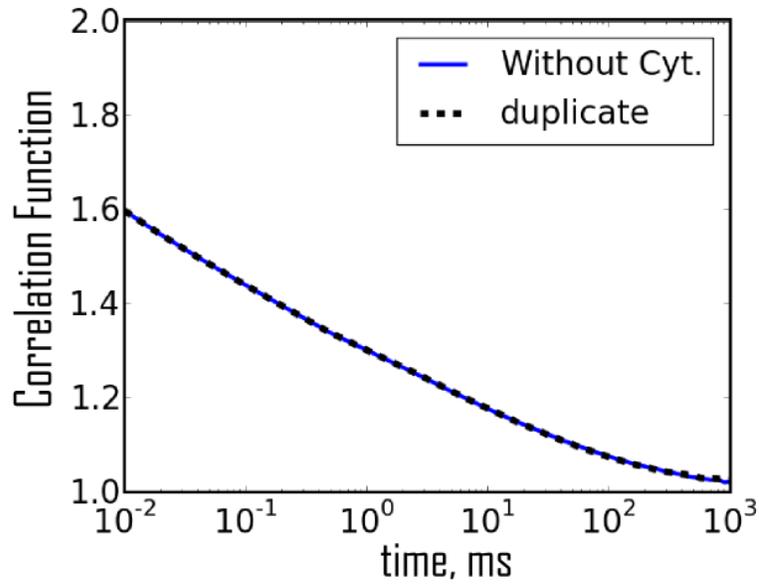

**Figure S4 To demonstrate the accuracy of our time-time correlation functions, we reproduce the slowest decaying curve from the fig. 4A and plot both versions on the same graph. A slight deviation is visible at very late times.**

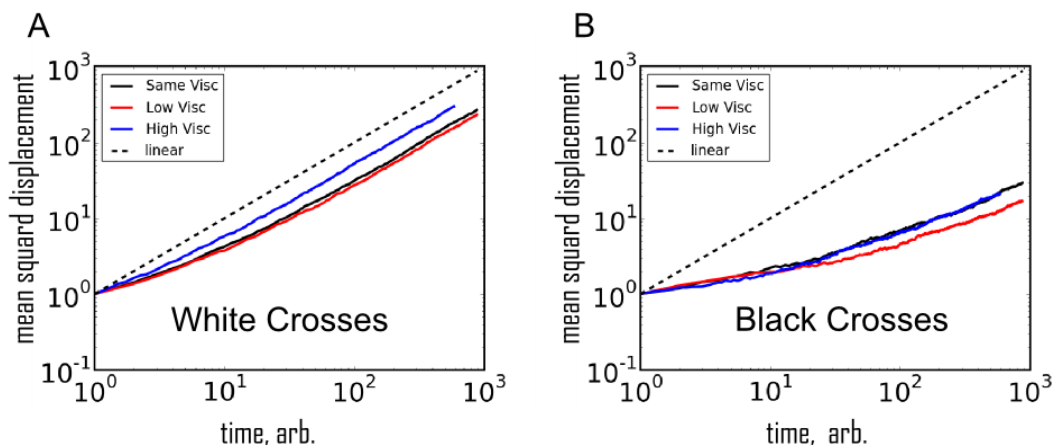

**Figure S5** The mean squared displacement of strongly coupled white (A) and black (B) diffusers at $1.05T_c$,with the same static properties as in Fig 4 D of the main text. In each figure the black line shows the mean square displacement for the case when both viscosities are equal. In the other cases the diffuser is a component of the high (blue) or low (red) viscosity liquid. The y-axis is normalized to 1 at a time such that the traced particle has performed on average 2000 attempted exchanges. The x-axis is normalized so that in each case the displayed traced particles are proposed to swap approximately 2000 times (0.5 ms in the main text) so that all have an identical microscopic attempt rate. As can be seen, particles that normally inhabit the low viscosity liquid see some degree of extra confinement, and vice versa.